\documentclass[a4paper,11pt]{article}
\pdfoutput=1 

\usepackage{jheppub} 

\usepackage[T1]{fontenc} 
\usepackage{amsmath,amssymb,euscript}
\usepackage{braket}
\usepackage{slashed}
\usepackage{xspace}
\usepackage{color}
\usepackage{hyperref}
\usepackage{epsfig}
\usepackage{xcolor}
\usepackage{xspace}
\usepackage{verbatim}
\usepackage{multirow}
\usepackage{booktabs,graphicx}
\usepackage{mathtools}
\usepackage{tabulary}
\usepackage{soul}
\usepackage{caption}
\usepackage{subcaption}
\usepackage{orcidlink}
\usepackage{xfrac}

\arxivnumber{2306.08686}
\preprint{DESY-23-088,\ Nikhef-2023-006}

\title{\boldmath Top Secrets: Long-Lived ALPs in Top Production}

\author[a,b]{L. Rygaard \orcidlink{0000-0003-3192-1622},}
\author[a]{J. Niedziela \orcidlink{0000-0002-9514-0799},}
\author[c]{R. Sch\"afer,}
\author[c,f]{S. Bruggisser,}
\author[a]{J. Alimena \orcidlink{0000-0001-6030-3191},}
\author[c,d,e]{S. Westhoff,}
\author[a,b]{F. Blekman \orcidlink{0000-0002-7366-7098}}  

\affiliation[a]{Deutsches Elektronen-Synchrotron DESY, Notkestr. 85, 22607 Hamburg, Germany} 
\affiliation[b]{Institut f\"ur Experimentalphysik, Universit\"at Hamburg, Luruper Chaussee 149, 22761 Hamburg, Germany} 
\affiliation[c]{Institute for Theoretical Physics, Heidelberg University, 69120 Heidelberg, Germany}
\affiliation[d]{Institute for Mathematics, Astrophysics and Particle Physics, Radboud University, 6500 GL Nij\-megen, The Netherlands}
\affiliation[e]{Nikhef, Science Park 105, 1098 XG Amsterdam, The Netherlands}
\affiliation[f]{Department of Physics and Astronomy, Uppsala University, 75120 Uppsala, Sweden}

\emailAdd{lovisa.rygaard@cern.ch}

\abstract{We investigate the discovery potential for long-lived particles produced in association with a top-antitop quark pair at the (High-Luminosity) LHC. Compared to inclusive searches for a displaced vertex, top-associated signals offer new trigger options and an extra handle to suppress background. We design a search strategy for a displaced di-muon vertex in the tracking detectors, in association with a reconstructed top-antitop pair.
 For axion-like particles with masses above the di-muon threshold, we find that the (High-Luminosity) LHC can probe effective top-quark couplings as small as $|c_{tt}|/f_a = 0.03~(0.002)/$TeV and proper decay lengths as long as $20~(300)$~m, assuming a cross section of $1\,$fb, with data corresponding to an integrated luminosity of 150~fb$^{-1}$ (3~ab$^{-1}$). Our predictions suggest that searches for top-associated displaced di-muons will explore new terrain in the current sensitivity gap between searches for prompt di-muons and missing energy.}

\begin{document} 
\maketitle
\flushbottom

\section{Introduction}
\label{sec:intro}
Collider searches for long-lived particles (LLPs)~\cite{Alimena:2019zri} explore the fourth space-time dimension where discoveries might hide: time. Particles with a long lifetime statistically decay at an observable distance from their production point. If the decay is inside the detector, their (charged) decay products can be reconstructed as a displaced vertex, a characteristic LLP signal.

Searches for displaced vertices bridge the sensitivity gap in decay length between searches for promptly decaying particles and searches with missing energy. Displaced vertex searches can even lead in sensitivity, especially if prompt and missing energy signals are plagued by a large Standard Model (SM) background. Exploiting the full reach of displaced vertex searches is key to maximizing the discovery potential for LLPs and to taking full advantage of colliders like the LHC.

At the LHC, currently, most searches for displaced vertices are either inclusive with respect to the remainder of the event~\cite{CMS:2022qej,CMS:2021sch,CMS:2021tkn,CMS:2020iwv,ATLAS:2018rjc,ATLAS:2019fwx,ATLAS:2023oti,ATLAS:2020xyo,ATLAS:2019tkk,ATLAS:2019qrr,LHCb:2017xxn} or target model-specific signatures of LLPs produced in Higgs decays~\cite{ATLAS:2018tup, ATLAS:2019jcm,ATLAS:2022gbw,ATLAS:2022zhj,ATLAS:2021jig,ATLAS:2018pvw,CMS:2021yhb,LHCb:2016buh}. However, inclusive searches often face trigger challenges~\cite{Alimena:2021mdu}, while model-specific searches are hard to reinterpret in other scenarios. LLP searches in association with other particles are a promising alternative, as demonstrated by recent pioneering searches~\cite{CMS:2023ovx,CMS:2022fut,ATLAS:2019kpx}. 

In this work, we explore the LHC's discovery potential for displaced vertices produced in association with prompt particles. We will call them \emph{associated displaced vertices} in what follows. Such signatures offer the option to trigger on the prompt part of the signal, which allows one to optimize the selection criteria for the displaced vertex. Compared to model-specific searches that rely on reconstructed event kinematics of the displaced vertex, associated displaced vertex searches can be kept fairly general by only applying loose kinematic selections on the prompt part of the signal. All that is needed is an efficient LLP production channel with a fully reconstructable final state. Examples are top-antitop production, the Drell-Yan process, Vector Boson Fusion, or di-jet production, which probe LLPs with couplings to top quarks, weak gauge bosons, and light quarks or gluons, respectively.

In this study, we focus on long-lived axion-like particles (ALPs, or ``$a$'') produced in association with a top-antitop pair, $pp\to t\bar{t}a$, and decaying to a displaced di-muon, $a\to~\mu^+\mu^-$. In general, such a signature is particularly sensitive to new light scalar or pseudo-scalar particles, which in many scenarios have flavor-hierarchical couplings to quarks~\cite{Froggatt:1978nt,Georgi:1986df,Gherghetta:2000qt,Agashe:2004rs}. In this paper, we therefore focus on ALPs with masses in the GeV range and only top couplings $|c_{tt}|/f_a$ at high energies, see e.g. Ref.~\cite{Bauer:2021mvw,Esser:2023fdo}. The decays of these and similar light top-philic particles are loop-suppressed, leading to a long decay length. They may thus have escaped searches for prompt resonances in association with top quarks~\cite{ATLAS:2023ofo,CMS:2019lwf,CMS:2022arx} and still not be long-lived enough to be caught in searches for tops and missing energy~\cite{CMS:2015zwg,CMS:2017dcx,CMS:2017qxu,CMS:2019zzl,CMS:2018ysw,ATLAS:2018cjd,ATLAS:2014bba,ATLAS:2018nda,ATLAS:2017hoo,ATLAS:2014dbf}. For ALPs with flavor-changing couplings, a complementary search for top-associated displaced ALP decays in the calorimeter has been recently put forward in Ref.~\cite{Carmona:2022jid}.

This article is structured as follows. In Sec.~\ref{sec:llps}, we discuss the basic characteristics of the signal and introduce the ALP benchmark model. In Sec.~\ref{sec:signal}, we present a detailed analysis of the signal and background kinematics and propose an efficient event selection. In Sec.~\ref{sec:predictions}, we predict the sensitivity to top-associated displaced di-muon events at the LHC during Run 2 and the High-Luminosity LHC (HL-LHC). We conclude in Sec.~\ref{sec:summary}.

\section{Long-lived particles in top-antitop production}
\label{sec:llps}
New scalars or pseudo-scalars are predicted in many extensions of the SM. Unlike new vector particles, scalars are generally expected to have flavor-hierarchical coup\-lings to quarks and leptons, with the strongest coupling to top quarks. The reason for this assumption is that scalar couplings to fermions break the flavor symmetry of the Lagrangian, just like the Higgs Yukawa couplings. To preserve the observed mass hierarchies and flavor mixing of the fermions, new scalar couplings should therefore not deviate much from the Yukawa-like flavor pattern in the SM.

At the LHC, top-antitop production is a natural place to search for new (pseudo-)scalar particles. If they couple mostly to top quarks, but are too light to decay into a top-antitop pair, their decay is suppressed. This suppression can be so strong that particles produced from top-antitop production decay with a significant displacement from their production point. In Fig.~1, we show a representative Feynman diagram of the LHC process 
\begin{align}\label{eq:signal-process}
pp \to t\bar{t} a,\, a\to \mu\bar{\mu}.
\end{align}
A resonant long-lived ALP is produced in top-antitop events from proton-proton collisions and decays into a displaced muon-antimuon pair. In general, other decay channels like $e^+ e^-$, $\tau^+\tau^-$, $\gamma\gamma$, or $q\bar{q}$ are possible, depending on the mass and couplings of the LLP. In this work, we focus on the $\mu^+\mu^-$ final state, which can be reconstructed well in the LHC detectors.


\begin{figure}[t!]
\centering
\includegraphics[width=.5\textwidth]{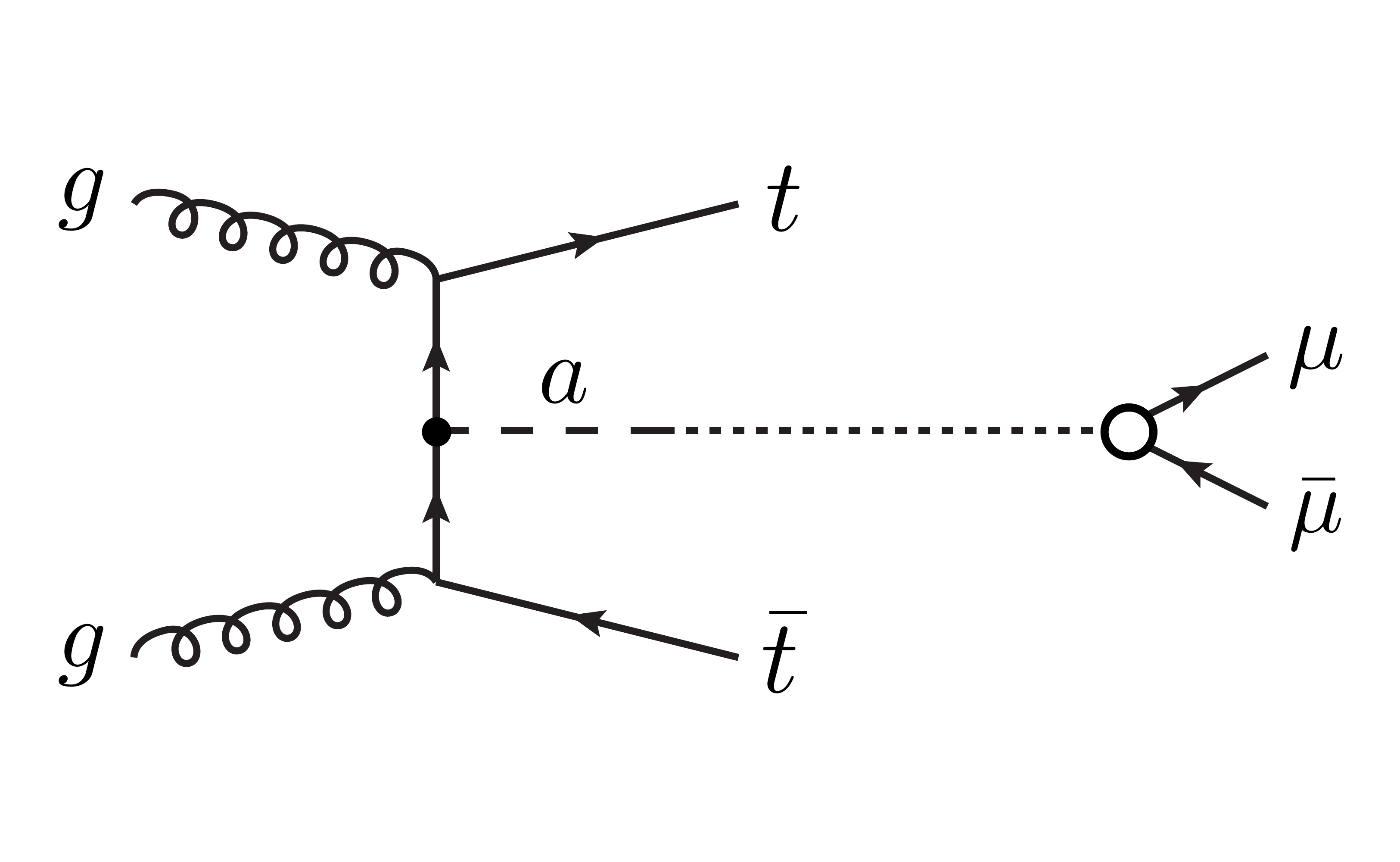}
\caption{\label{fig:feynman}Example Feynman diagram for the production and decay of a long-lived ALP at the LHC via $pp \to t\bar{t} a,\, a\to \mu\bar{\mu}$. The dotted line illustrates that ALP decays at a distance from its production point. The small black circle indicates the effective ALP-top coupling, and the large white circle indicates the loop-induced ALP coupling to muons.}
\end{figure}

\subsection{Signal event rate}\label{sec:rate}
The expected number of top-associated LLPs ``$a$'' produced at the LHC with a given integrated luminosity $\mathcal{L}$ is
\begin{align}
N = \mathcal{L}\,\sigma(pp\to t\bar{t}a),
\end{align}
where $\sigma$ is the hadronic cross section at a fixed proton-proton collision energy. The decay length of the LLP in the laboratory frame, $d_a = \beta\gamma c\tau_a$, is determined by its lifetime $\tau_a$ and Lorentz boost $\beta\gamma$. For an LLP with decay length $d_a$, produced in the direction $\vec{r}_a$, the probability to decay within a radial interval $\Delta r_a = r_a^{\rm out} - r_a^{\rm in}$ is
\begin{align}\label{eq:decay-prob}
P_a(d_a; \vec{r}_a) = \exp\left(-\frac{r_a^{\rm in}}{d_a}\right) - \exp\left(-\frac{r_a^{\rm out}}{d_a}\right).
\end{align}
In practice, $r_a^{\rm in}$ and $r_a^{\rm out}$ are the distances from the production point to where the particle's trajectory intersects with the detector boundaries. The decay distribution of a sample of particles is thus exponential, with the slope determined by the proper lifetime $\tau$ and the average boost of the particles. On the other hand, displacement is a property of each particle. It refers to where in the detector that particle decays. We will refer to particles that decay after reaching at least 200 $\mu$m as \emph{displaced}, and as \emph{prompt} otherwise.

For a sample of $N$ produced LLP events, the fraction of LLPs that decay within a detector volume $V$ is
\begin{align}\label{eq:na}
N_a(V) = \mathcal{L}\, \frac{1}{N}\sum_{a=1}^N \sigma\left(pp \to t\bar{t}\, a\,|d_a;\vec{r}_a\right) P_a\left(d_a;\vec{r}_a\right),
\end{align}
where $\sigma\left(pp \to t\bar{t}\, a\,|d_a;\vec{r}_a\right)$ is the cross section for an LLP produced in a small interval around the direction $\vec{r}_a$ with boost $\beta\gamma = d_a/c\tau_a$.
The expected number of $t\bar{t}(\mu\bar{\mu})$ signal events is finally given by
\begin{align}
N_{\rm sig} = N_a(V)\,\mathcal{B}(a\to \mu\bar{\mu}),\qquad \mathcal{B}(a\to \mu\bar{\mu}) = \frac{\Gamma(a\to \mu\bar{\mu})}{\Gamma_a},
\end{align}
where $\Gamma_a = \tau_a^{-1}$ and $\Gamma(a\to \mu\bar{\mu})$ are the total decay width and the partial decay width of the LLP into muons and $\mathcal{B}(a\to \mu\bar{\mu})$ is the LLP's branching ratio into muons.

In general, the signal rate $N_{\rm sig}$ depends on four fundamental properties of the LLP: the mass $m_a$, the coupling to top quarks $c_{tt}$, the decay width $\Gamma_a$, and the branching ratio $\mathcal{B}(a\to \mu\bar{\mu})$. In specific models, some of these quantities might be related.

\subsection{Axion-like particles}\label{sec:alps} ALPs are pseudo-scalars whose couplings preserve a shift symmetry $a\to a + c$. At energies below a cutoff scale $\Lambda = 4\pi f_a$, the phenomenology is described by an effective Lagrangian~\cite{Georgi:1986df}
\begin{align}\label{eq:lagrangian}
\mathcal{L}_{\rm eff}(\mu) & = \frac{1}{2}(\partial_\mu a)(\partial^\mu a) - \frac{m_a^2}{2}a^2 + \sum_{V}c_{VV}(\mu)\,\frac{a}{f_a}\frac{g^2}{(4\pi)^2}\, V_{\mu\nu}\widetilde{V}^{\mu\nu}\\\nonumber
& \quad + \sum_{F=Q,L}\frac{\partial^\mu a}{f_a}\,\bar{F}_i\,(C_F)_{ij}\,\gamma_\mu \,F_j + \sum_{f=U,D,E}\frac{\partial^\mu a}{f_a}\,\bar{f}_i\,(C_f)_{ij}\,\gamma_\mu \,f_j.
\end{align}
In general, the scale $f_a$ and the mass of the ALP, $m_a$, are free parameters.\footnote{In the special case of the QCD axion, $f_a$ corresponds to the axion's decay constant, and the axion mass scales as $m_a \propto 1/f_a$.} In our analysis, we set $f_a = 1\,$TeV. The parameters $c_{VV}$ label ALP couplings to gauge fields $V = G,W,B$ associated with the $SU(3)_C$, $SU(2)_L$ and $U(1)_Y$ groups, with field-strength tensor $V_{\mu\nu}$ and dual $\widetilde{V}_{\mu\nu} = \epsilon_{\mu\nu\rho\sigma}V^{\rho\sigma}$. The ALP couplings $C_F$ and $C_f$ to left-handed and right-handed fermions, respectively, are in general $3\times 3$ matrices in flavor space.\footnote{The $SU(2)_L$ doublets are defined as $Q=(u_L,d_L)^\top$ and $L = (\nu_L,\ell_L)^\top$; the $SU(2)_L$ singlets are $U=u_R$, $D=d_R$, $E = \ell_R$.} Among the flavor-diagonal couplings, only the axial-vector combinations $(C_F)_{ii}-(C_f)_{ii}$ are physical; vector couplings are incompatible with the shift symmetry. The derivative $\partial^\mu a$ implies that ALP couplings to on-shell fermions are proportional to the fermion mass $m_f$ and therefore generally flavor-hierarchical.

We choose to work in a basis where the interaction and mass eigenstates of up-type quarks coincide. In this basis, we define the flavor-diagonal ALP couplings to up-type and down-type quarks as well as to charged leptons as
\begin{align}
    c_{ii} & = (C_u)_{ii} - (C_Q)_{ii},\qquad\quad\ \, i = \{u,c,t\}\\\nonumber
    c_{jj} & = (C_d)_{jj} - (V^\dagger C_Q V)_{jj},\quad j = \{d,s,b\}\\\nonumber
    c_{kk} & = (C_\ell)_{kk} - (C_L)_{kk},\qquad\ \ \, k = \{e,\mu,\tau\},
\end{align}
where $V$ is the CKM matrix. Flavor off-diagonal couplings will not play a role in this work.

The top-quark coupling $c_{tt} = (C_u)_{33} - (C_Q)_{33}$ plays an important role in the ALP effective theory at low energies. In the renormalization group (RG) evolution of the effective couplings, $c_{tt}$ generates ALP couplings to all other fermions through mixing and renormali\-za\-tion effects~\cite{Choi:2017gpf,MartinCamalich:2020dfe,Chala:2020wvs,Bauer:2020jbp}. For $\Lambda = 4\pi\,$TeV, the ALP couplings at the top mass scale $\mu = m_t$ are
\begin{align}\label{eq:running}
c_{uu,cc}(m_t) & = c_{uu,cc}(\Lambda) - 0.116\, c_{tt}(\Lambda)\\\nonumber
c_{dd,ss}(m_t) & = c_{dd,ss}(\Lambda) + 0.116\, c_{tt}(\Lambda)\\\nonumber c_{bb}(m_t) & =  c_{bb}(\Lambda) + 0.097\, c_{tt}(\Lambda)\\\nonumber
c_{\ell\ell}(m_t) & = c_{\ell\ell}(\Lambda) + 0.116\, c_{tt}(\Lambda),\quad \ell = \{e,\mu,\tau\}.
\end{align}
To a good approximation, these relations also hold in the perturbative regime below the weak scale. The ALP-top coupling can thus be probed in low-energy observables through RG effects in effective ALP couplings to light fermions. Moreover, the ALP-top coupling induces ALP couplings to photons and gluons through top loops, which however play a minor role in this work. Notice that also the top coupling itself receives RG corrections at energies below the cutoff scale. For numerical predictions of the ALP couplings at different energy scales, we use our public code~\cite{Schafer_UFO} that implements the results of Ref.~\cite{Bauer:2020jbp}.

A very predictive scenario is obtained by assuming that the top coupling is the only coupling of the ALP at the cutoff scale, so that
\begin{align}\label{eq:top-only}
c_{tt}(\Lambda) \equiv c_{tt},\qquad c_{f\!f}(\Lambda) = 0\ (f \neq t),\qquad c_{VV}(\Lambda) = 0.
\end{align}
We will call this scenario the~\emph{top scenario} in what follows.
Such a scenario arises from ultraviolet completions of the ALP effective theory with top-philic heavy particles that generate the effective coupling $c_{tt}(\Lambda)$. In the top scenario, all ALP couplings are proportional to $|c_{tt}|/f_a$. As a consequence, all partial decay widths of the ALP are determined by $|c_{tt}|/f_a$. Moreover, the ALP decay rate is fully correlated with the production from top quarks. The signal rate is thus determined by only two parameters: the ALP mass $m_a$ and the effective top coupling $|c_{tt}|/f_a$.

The total decay width of the ALP is in general given by~\footnote{Decays into neutrinos are in principle possible, but extremely suppressed by the small neutrino mass.}
\begin{align}
\Gamma_a = \Gamma(a \to \gamma\gamma) + \sum_{\ell = e,\mu,\tau} \Gamma(a \to \ell\bar{\ell})\, \Theta(m_a - 2 m_\ell) + \Gamma(a \to \text{had})\, \Theta(m_a - 3 m_\pi).
\end{align}
The ALP decay into leptons and hadrons is open only above the kinematic threshold, as indicated by the Heaviside function $\Theta$. The decay rate into fermion pairs is
\begin{align}\label{eq:decay-rate-fermions}
\Gamma(a\to f\!\bar{f}) =  \frac{m_a}{8\pi} \frac{\vert c_{f\!f}(m_a)\vert^2 m_f^2}{f_a^2}\left(1-\frac{4m_f^2}{m_a^2}\right)^{\frac{1}{2}}.
\end{align}
In the top scenario, $c_{f\!f}(m_a) \approx c_{f\!f}(\mu_w)$ for ALP masses below the weak scale $\mu_w$, and Eq.~\ref{eq:running} applies to a good approxi\-mation. For hadronic decays, Eq.~\ref{eq:decay-rate-fermions} applies only in the regime of perturbative QCD, \emph{i.e.}, for $m_a \gtrsim 2\,$GeV. To make predictions at lower scales, the ALP effective theory has to be matched onto the chiral perturbation theory~\cite{Georgi:1986df,Bauer:2021wjo}.

\begin{figure}[t!]
\centering
\includegraphics[width=.495\textwidth]{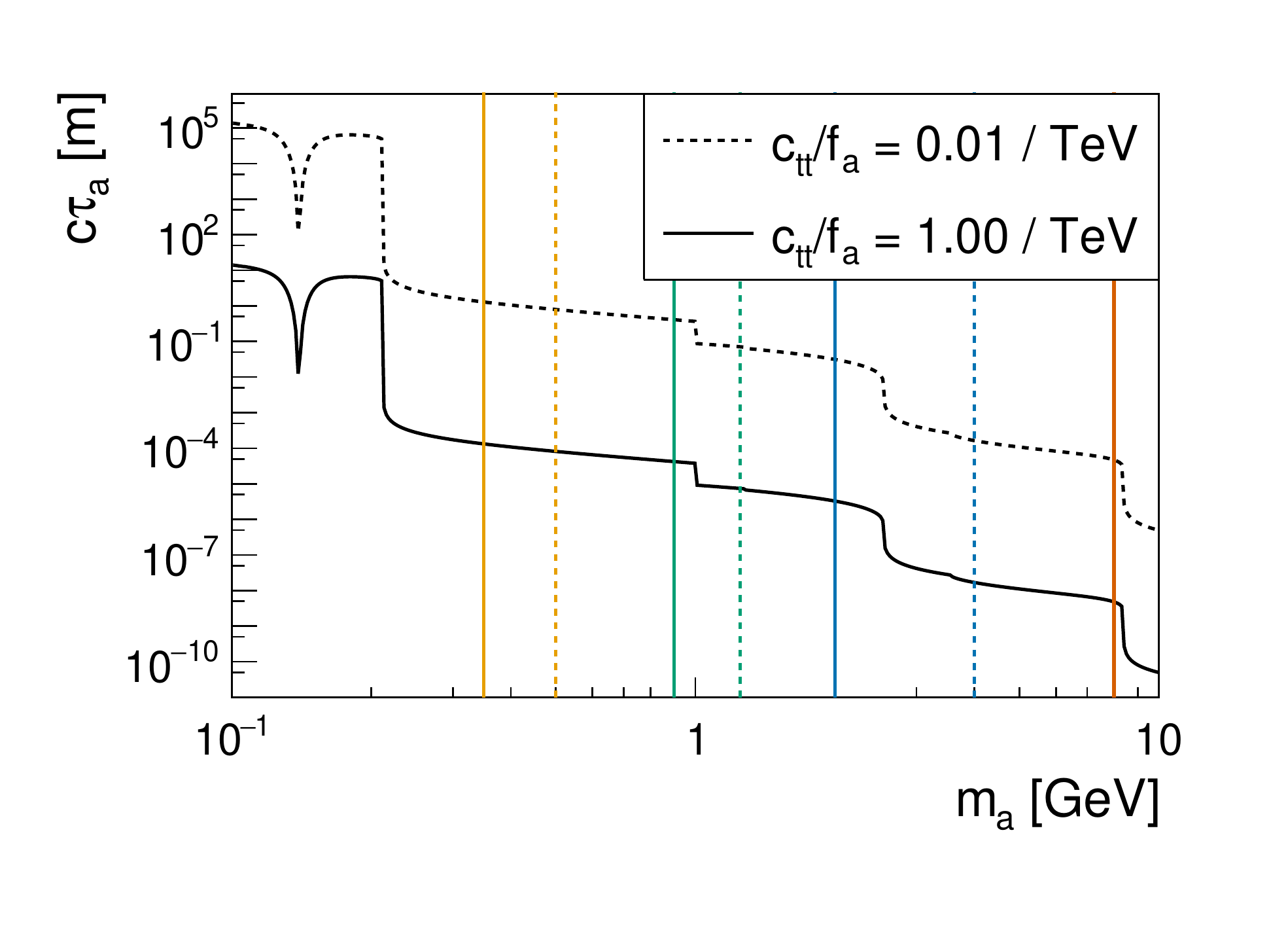}
\includegraphics[width=.495\textwidth]{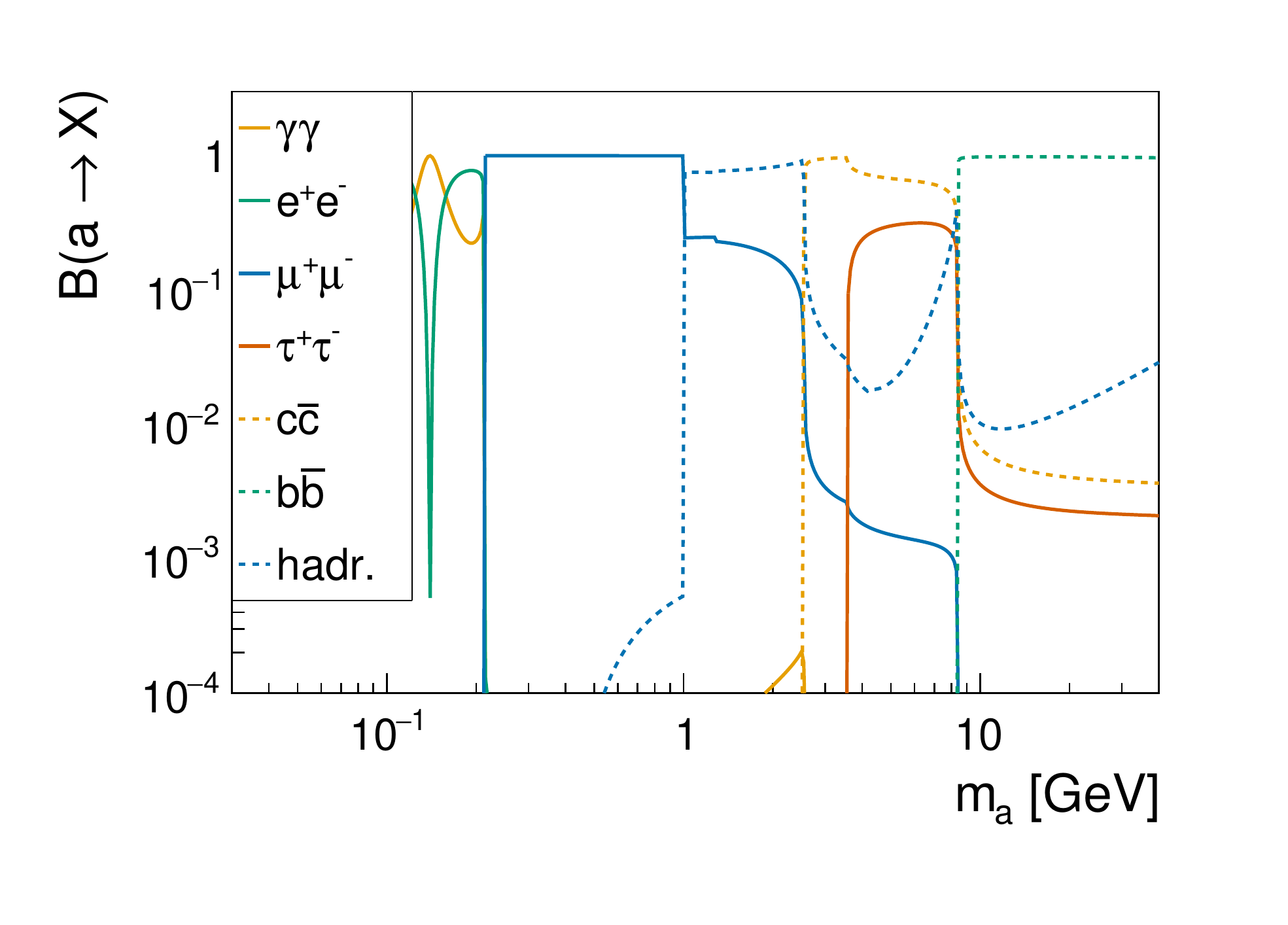}
\caption{\label{fig:decay-length}Proper decay length, $c\tau_{a}$, (left) and branching ratios (right) of an ALP in the top scenario (Eq.~\ref{eq:top-only}) as a function of its mass, $m_a$. For a fixed mass, all decay modes are fully determined by the top-quark coupling $|c_{tt}(\Lambda)|/f_a$, defined at the cutoff scale $\Lambda = 4\pi\,$TeV. In the left panel, vertical lines for $c\tau_{a}$ indicate benchmark ALP masses used in this study. The dip in the decay length around $m_a = m_\pi$ is due to ALP-pion mixing.}
\end{figure}

In Fig.~\ref{fig:decay-length} (left), we show the proper decay length of the ALP in the top scenario (Eq.~\ref{eq:top-only}) as a function of its mass. Since all partial decay rates are determined by the top coupling, the decay length scales as $c\tau_a \propto f_a^2/c_{tt}^2$. The ALP branching ratio into any specific final state is thus independent of $c_{tt}$. The top scenario (Eq.~\ref{eq:top-only}) predicts the longest possible decay length of ALPs produced from top quarks. Any additional coupling would increase the decay width, resulting in a shorter decay length.

The branching ratios of ALP decays in the top scenario are shown in Fig.~\ref{fig:decay-length} (right), as a function of the ALP mass. For $m_a < 2 m_\mu$, the ALP can only decay into photons and, if kinematically allowed, into electrons. For $2 m_\mu < m_a < 1\,\text{GeV}$, the ALP width is dominated by the decay into muons; decays into electrons and photons are suppressed by the small electron mass and the electromagnetic coupling, respectively. For ALP masses around the GeV scale, predictions are affected by large uncertainties in hadronic decays. For the exact implementation of the ALP decay channels, we refer the reader to App.~A of Ref.~\cite{Ferber:2022rsf}.

Where perturbative predictions are possible, the ALP branching ratio into muons in the top scenario is given by
\begin{align}
2 m_\mu < m_a < 1\,\text{GeV}: \qquad \mathcal{B}(a \to \mu\bar{\mu}) & \approx 1\\\nonumber
m_a > 2\,\text{GeV}: \qquad \mathcal{B}(a \to \mu\bar{\mu}) & \approx \frac{\Gamma(a \to \mu\bar{\mu})}{\sum_f\Gamma(a\to f\!\bar{f})\,\Theta(m_a - 2 m_f)}\,.
\end{align}
Additional couplings beyond $c_{tt}$ can significantly change the branching ratios.

For our analysis of displaced vertices, we focus on ALPs within the mass range
$2m_\mu < m_a < 2 m_b$. Below the di-muon threshold, the decay into muons is kinematically forbidden; above the $b\bar{b}$ threshold, decays into bottom-antibottom pairs result in essentially prompt decays. Within the mass range, we define the following benchmarks:
\begin{align}\label{eq:benchmarks}
m_a = \{0.3,\,0.35,\,0.5,\,0.9,\,1.25,\,2,\,2.5,\,2.75,\,3,\,4,\,8,\,8.25,\,8.5,\,9,\,10\}\,\text{GeV}.
\end{align}
In the first two benchmarks, the ALP decays 100\% of the time into muons. In all other benchmarks, additional decay channels are kinematically allowed and reduce the branching ratio to muons. The impact of the various decay channels on the proper decay length is shown in Fig.~\ref{fig:decay-length}. We will use these benchmarks to study the dependence of the signal features on the ALP mass and lifetime.

When comparing these benchmarks in the top scenario with concrete UV completions of the ALP effective theory, it is important to know for which scenarios they are good proxies. Additional ALP couplings at the cutoff scale affect the lifetime and branching ratios of the ALP. In Sec.~\ref{sec:general-results}, we therefore interpret our results for ALPs with arbitrary lifetimes. Additional ALP couplings can also lead to new production channels. In the quark sector, the phenomenology is usually still dominated by the top coupling, because ALP couplings to on-shell fermions are proportional to the fermion mass. Besides resonant top production, sensitive observables are, for instance, flavor-changing $B$ meson and kaon decays, which probe the top coupling in loops. In the lepton sector, the mass proportionality can be outrun by precision probes of the ALP-electron coupling. To obtain maximal information about the ALP couplings in a specific UV completion, a global fit of the ALP effective Lagrangian to data is the way to go~\cite{Bruggisser:2023npd,Biekotter:2023mpd}.

\section{Signal and background features}
\label{sec:signal}
The goal of this analysis is to determine the LHC sensitivity to a long-lived pseudo-scalar particle produced in association with a top-antitop pair, as in Eq.~\ref{eq:signal-process}. We perform a~dedicated analysis of the signal and background kinematics for the ALP benchmarks from Eq.~\ref{eq:benchmarks} in the top scenario, Eq.~\ref{eq:top-only}. The results are then generalized to ALPs with an arbitrary lifetime.

In the top scenario, the phenomenology of the ALP is determined by two parameters, the mass $m_a$ and the top-quark coupling $|c_{tt}(\Lambda)|/f_a$. Throughout the analysis, we set $f_a~=~1\,$TeV. In our analysis of kinematics and decay distributions, we fix $c_{tt}(\Lambda)~=~1$, such that the signal rate only depends on the ALP mass.

The main SM background processes that can mimic a di-muon vertex in association with a top-antitop pair are:
\begin{itemize}
\item $pp \to t\bar{t}Z^{(\ast)},\ Z^{(\ast)}\to \mu\bar{\mu}$: top-antitop production in association with a virtual or resonant $Z$ boson (or photon) decaying/converting into a di-muon;
\item $pp \to t\bar{t}j$: top-antitop production in association with one or more jets, where hadrons inside the jet decay into two opposite-sign muons.
\end{itemize}
The $t\bar{t}Z^{(\ast)}$ process is an important background for prompt or nearly prompt di-muons. For top-associated displaced vertices, $t\bar{t}Z^{(\ast)}$ can still be a background because the parton shower produces mesons with a significant lifetime that decay into displaced di-muons. The $t\bar{t}j$ background is relevant at larger displacements because the muons are produced from decays of mesons inside the jet with a significant boost. In both cases, di-muons can originate from the same particle (e.g. a $J/\Psi$ meson), which we will refer to as "resonant". On the other hand, di-muons can come from decays of two different mesons, called "non-resonant" in what follows.

Muons from cosmic rays produce displaced muons in the LHC detectors, and therefore in principle, cosmic muons that randomly cross and produce a displaced vertex could be a background for this analysis. However, the rate of this process is small and in practice, there are many ways for the experiments to reject the vast majority of this background~\cite{CMS:2022qej,CMS:2021sch,ATLAS:2018rjc,ATLAS:2019fwx}. As a result, we do not consider background from cosmic muons in this analysis.

We focus on the extraction of the displaced di-muon vertex and assume a perfect reconstruction of the top-antitop pair (in all possible final states) and a 100\% effective top trigger. We also assume perfect identification of muons coming from top quark decays, meaning not only muons originating from the W boson decay, but also those originating from the hadronization of the b quark. We estimate the impact of this assumption on the sensitivity to LLPs in Sec.~\ref{sec:predictions}.
\subsection{Event simulation}
All signal and background processes are generated using \texttt{MadGraph5\_aMC@NLO} v. 3.2.0~\cite{Alwall:2014} to simulate proton-proton collisions at $\sqrt{s} = 13\,$TeV, with \texttt{NNPDF31\_NLO\_as\_0118} parton distribution functions \cite{Ball_2017}. All simulations of the hard scattering process are performed at leading order in QCD. Hadronization and particle showering are simulated using \texttt{PYTHIA8} v.8.306 \cite{PYTHIA8}, with the default Monash 2013 tune \cite{Skands_2014}. The top mass is set to $m_t = 172.0\,$GeV.

For the generation of the signal process $pp\to t\bar{t}a$, we use a UFO model~\cite{Schafer_UFO} that implements the effective Lagrangian (Eq.~\ref{eq:lagrangian}) in the top scenario. We set the ALP-top coupling to $|c_{tt}(\Lambda)|/f_a = 1$/TeV and all other ALP couplings to zero. The displaced ALP decay is simulated by \texttt{PYTHIA8} based on the decay width, as described in Sec.~\ref{sec:rate} and~\ref{sec:alps}. ALPs with decays to hadrons or tau leptons can produce secondary muon pairs from meson or tau decays. For the tau channel, this occurs in less than 3 percent of all events. We~ignore such secondary muons in our study; their impact on our results is negligible.

For the background process $pp \to t\bar{t}Z^{(\ast)},\ Z^{(\ast)}\to \mu\bar{\mu}$, we require the muons to have a~transverse momentum $p_T^\mu > 5\,$GeV and pseudo-rapidity $|\eta^\mu| < 3.5$ at generator level. 

For~$pp \to t\bar{t} j$, we first generate $pp \rightarrow t \bar{t} j$ in \texttt{MadGraph5}, requiring that the jet has $p_T^j > 20\,$GeV and $|\eta^j| < 3$. These criteria are chosen such that the particles can be well reconstructed at the LHC detectors while maintaining optimal use of CPU resources by not generating events that do not fall within the typical kinematic acceptance of the LHC. Then we use \texttt{PYTHIA8} to add additional ISR/FSR jets. However, the cross section we obtain from \texttt{MadGraph5} only includes the $pp \to t\bar{t}+1j$ process. To include more jets we use the measured cross sections for the $t \bar{t}$ process with a different number of additional jets (0-4) from CMS~\cite{CMS:2017ttbar}. Since the kinematic selections in this analysis somewhat differ from ours and since CMS only included the semi-leptonic decay of the top quark, we use the ratio of multi-jet versus one-jet emission from CMS to re-scale our one-jet cross section obtained from \texttt{MadGraph5} (MG). Finally, our prediction of the $t\bar{t} + (1-4)j$ cross section is
\begin{align}
\sigma_{pp \to t\bar{t}+(1-4)j} = \sigma^{\mathrm{MG}}_{pp \to t\bar{t}+1j} \cdot \frac{\sigma^{\mathrm{CMS}}_{pp \to t\bar{t}+(1-4)j}}{\sigma^{\mathrm{CMS}}_{pp \to t\bar{t}+1j}} = 395~\mathrm{pb} \cdot \frac{132~\mathrm{pb}}{77~\mathrm{pb}} = 667~\mathrm{pb}.
\end{align}
\subsection{Event selection}\label{sec:selection}
We apply selection criteria to the event samples in two stages: pre-selection and signal selection. In the pre-selection, we select events that contain a displaced di-muon; in the signal selection, we subsequently apply specific requirements on the pre-selected di-muons to increase the signal sensitivity. Although we apply selections on the truth level objects, the selection requirements were tuned based on known detector resolutions, using the current CMS detector as a reference. A summary of the selection criteria is given in Table~\ref{tab:selection-summary}. 
\begin{table}[tbp]
    \centering 
    \resizebox{0.55\textwidth}{!}{ 
    \begin{tabular}{l l}
        \hline
        \hline
        Pre-selection &  \\ \hline
        Muon kinematics &  $p_T^\mu > 5\,$GeV,$\ |\eta^\mu| < 2.5$ \\
        Muon displacement & $l_{xy} > 200\,\mu$m \\
        \multicolumn{2}{c}{At least one opposite-sign di-muon} \\
         \hline
         \hline
        Signal selection & \\ \hline
        Muon kinematics & $p_T^\mu > 10\,$GeV \\
        Di-muon mass & $m_{\mu\bar{\mu}} \neq m_{J/\Psi}, m_{\Psi(2S)}$ \\
        Di-muon vertex & $R_{lxy} < 0.05$ \\
        \hline 
        \hline
    \end{tabular}} 
    \caption{\label{tab:selection-summary}Selection criteria applied on the signal and background events.}
\end{table}

\subsubsection{Pre-selection} 
\label{sec:preselection}

In the pre-selection, we request two opposite-sign muons, each passing basic kinematic selections $|\eta^\mu| < 2.5$ and $p_T^\mu > 5\,$GeV. We only consider muons that do not originate from $W$~bosons from top quark decays. Due to the re-balancing of momentum during hadronization in \texttt{PYTHIA}, muons coming from ALP decays may experience a slight modification of their four-momenta and be included multiple times in the event with different status codes. The modification in the four-momenta results in a di-muon mass spectrum with a non-negligible width, as will be shown below. Furthermore, to avoid non-physical distortions of the ALP resonance caused by this effect, for the signal we select the muon four-momentum before these modifications occur. For the background, we work with final-state muons.

The main criterion to identify a displaced muon is the transverse displacement $l_{xy}$, defined in terms of the position coordinates $x$ and $y$ of the muon production point as 
\begin{align}\label{eq:lxy}
l_{xy} = \sqrt{x^2 + y^2}.
\end{align}
In our analysis, we select displaced muons with $l_{xy} > 200\,\mu$m. This value is roughly based on the transverse impact parameter resolution at CMS (20--75~$\mu$m for tracks with $1\,\text{GeV} < p_T < 10\,$GeV~\cite{CMS:2014pgm}). We choose the conservative value of 200~$\mu$m.

This selection criterion significantly reduces the prompt $t\bar{t}Z$ background, with events passing mainly due to meson decays within jets, similarly to the $t\bar{t}j$ background. Given the small cross section of $t\bar{t}Z$, the expected number of passing events is negligible. Consequently, we do not consider this process any further in our analysis.

The event passes the pre-selection if there is at least one opposite-sign di-muon remaining after the above-mentioned criteria.

\subsubsection{Signal selection}\label{sec:signal-selection} In the second stage, we apply additional selection criteria that enhance the signal sensitivity. These criteria are motivated by the kinematic distributions of the signal and background processes. In what follows, we discuss the event kinematics in detail. The event rates displayed in the figures correspond to an integrated luminosity of 150 fb$^{-1}$.

\begin{figure}[t!]
\centering
\includegraphics[width=.85\textwidth]{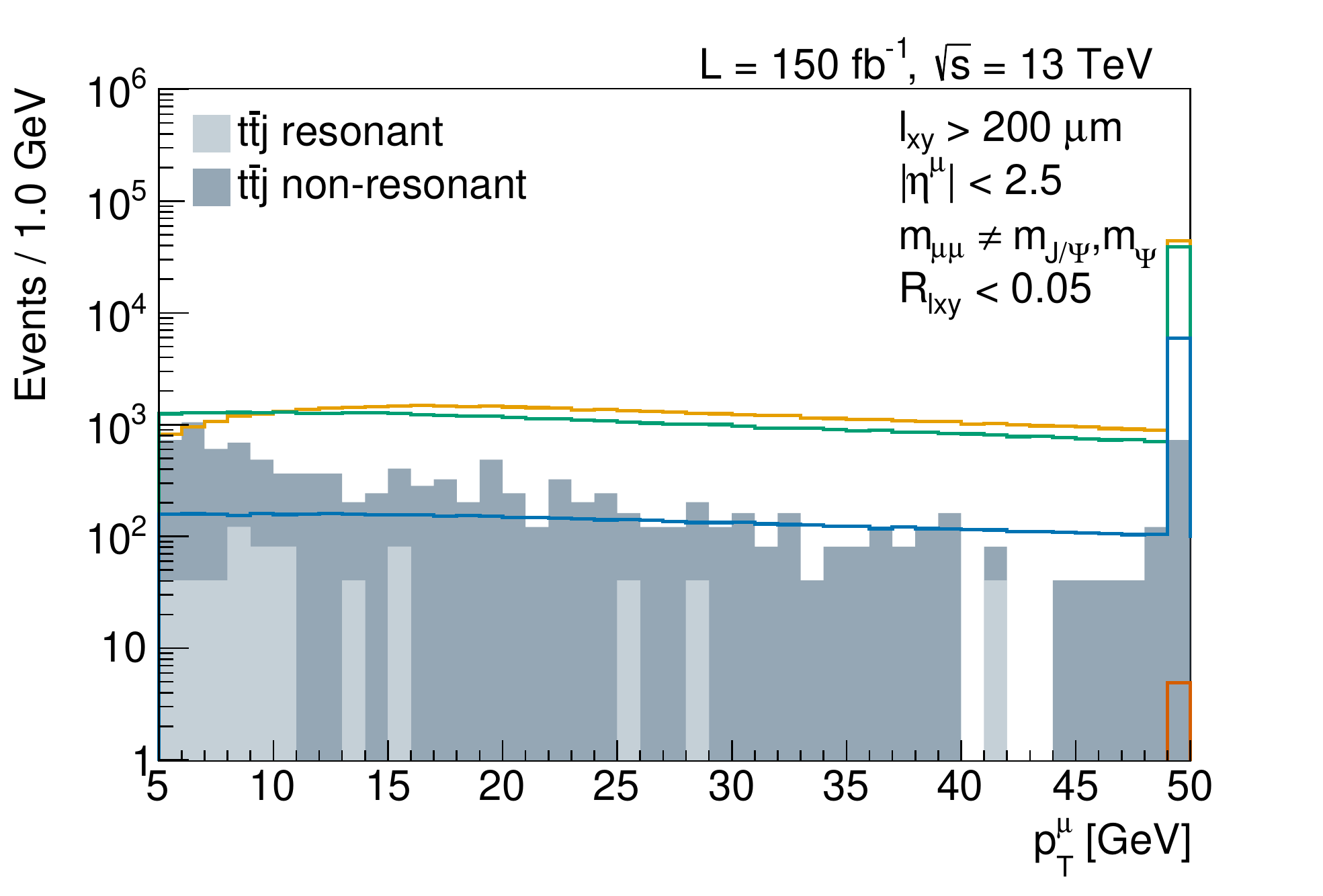}
\caption{\label{fig:muon-pt}The transverse momentum distribution of the muon with the lowest $p_T^\mu$ in each event, for various ALP signal mass benchmarks (colored lines) and background from $t\bar{t}j$ events (gray regions). The distribution is shown after the full event selection (upper right corner), excluding the $p_T^\mu$ criterion. The last bin includes the overflow at higher $p_T^\mu$.}
\end{figure}

Figure~\ref{fig:muon-pt} shows the transverse momentum distribution of the muon with the lowest $p_T^\mu$ in each event. The signal event rate depends little on the muon transverse momentum, with a slight decline for light ALPs at low $p_T^\mu$. The $t\bar{t}j$ background rate decreases at high $p_T^\mu$. To increase the signal sensitivity, we therefore tighten the selection of the muon transverse momentum to $p_T^\mu > 10\,$GeV.

\begin{figure}[t!]
\centering
\includegraphics[width=.85\textwidth]{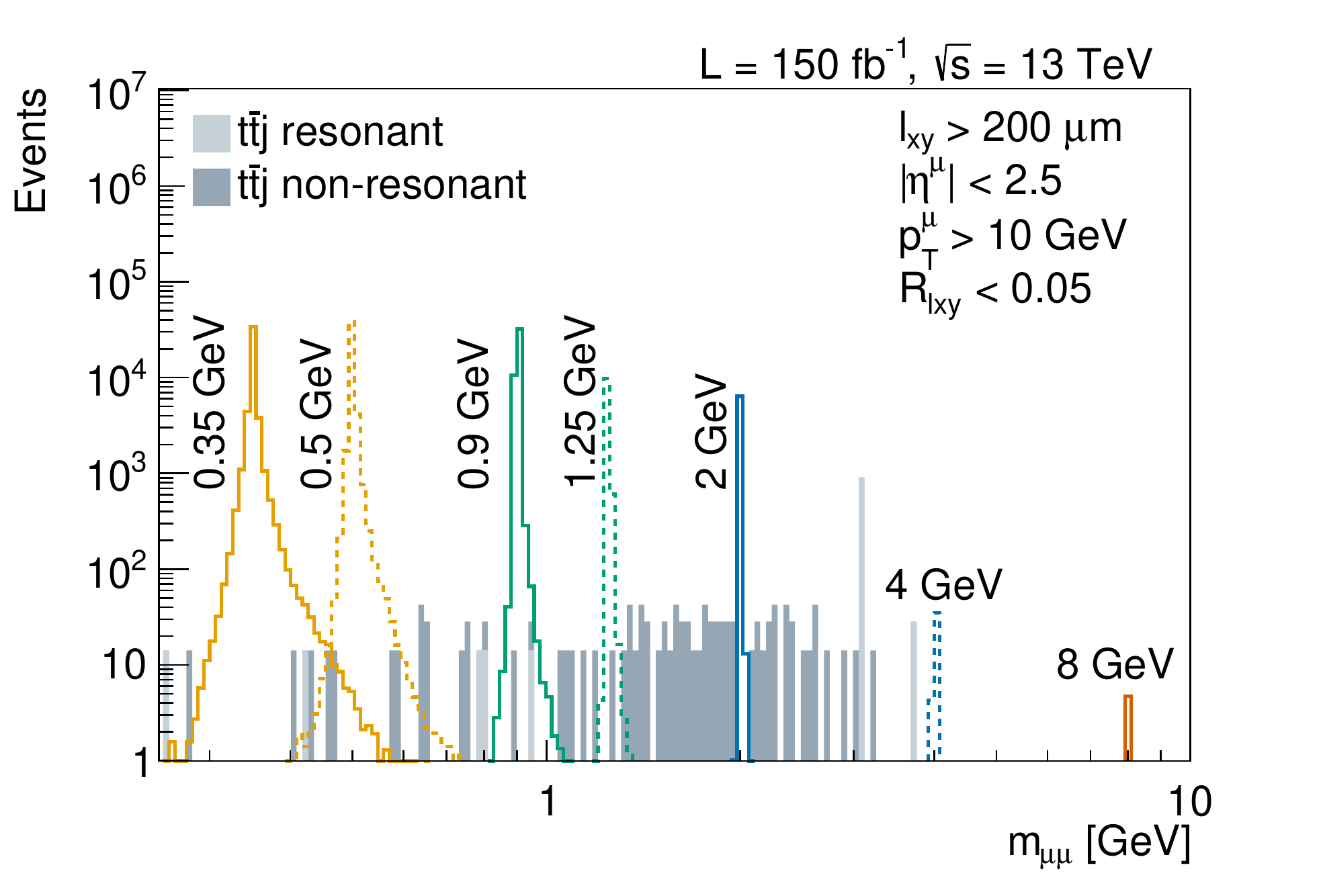}
\caption{\label{fig:di-muon-mass}The invariant mass distribution of the pre-selected di-muon in each event, for various ALP signal mass benchmarks (colored lines) and background from $t\bar{t}j$ events (gray regions). The distribution is shown after the full event selection (upper right corner), excluding the $m_{\mu\mu}$ criterion.}
\end{figure}

The di-muon invariant mass is shown in Fig.~\ref{fig:di-muon-mass}. The signal peaks around the ALP mass, as expected. The observable width for light ALPs is due to the four-momenta modifications in \texttt{PYTHIA} described in Sec.~\ref{sec:preselection}. The decrease in the number of signal events for higher ALP masses is mostly due to the lower branching ratio $\mathcal{B}(a\to \mu\bar{\mu})$. The peaks in the background are identified as the resonances of the $J/\Psi$ and $\Psi$(2S) mesons. We reject events with di-muon invariant masses within 5\% above and below the tabulated PDG masses of these mesons~\cite{Workman:2022ynf}. 

In experiments, the displaced vertex is reconstructed from the decay products. However, in this phenomenological study, we determine if two displaced muons originate from the same vertex by calculating the proximity of their production points. As a selection criterion, we use the difference in transverse displacement between the two muons. To reduce the dependency on detector resolution assumptions, we define the ratio
\begin{align}\label{eq:rlxy}
R_{lxy} = \frac{ \sqrt{ (x_\mu - x_{\bar{\mu}})^2 + (y_\mu - y_{\bar{\mu}})^2}} {\sqrt{ (|x_\mu| + |x_{\bar{\mu}}|)^2 + (|y_\mu| + |y_{\bar{\mu}}|)^2}}
\end{align}
for the transverse coordinates of the muon, $x_\mu$ and $y_\mu$, and antimuon, $x_{\bar{\mu}}$ and $y_{\bar{\mu}}$. To ensure a smooth behavior for $x_\mu = - x_{\bar{\mu}},\, y_\mu = - y_{\bar{\mu}}$, we use the absolute values in the denominator of $R_{lxy}$.
 
\begin{figure}[t!]
\centering
\includegraphics[width=0.85\linewidth]{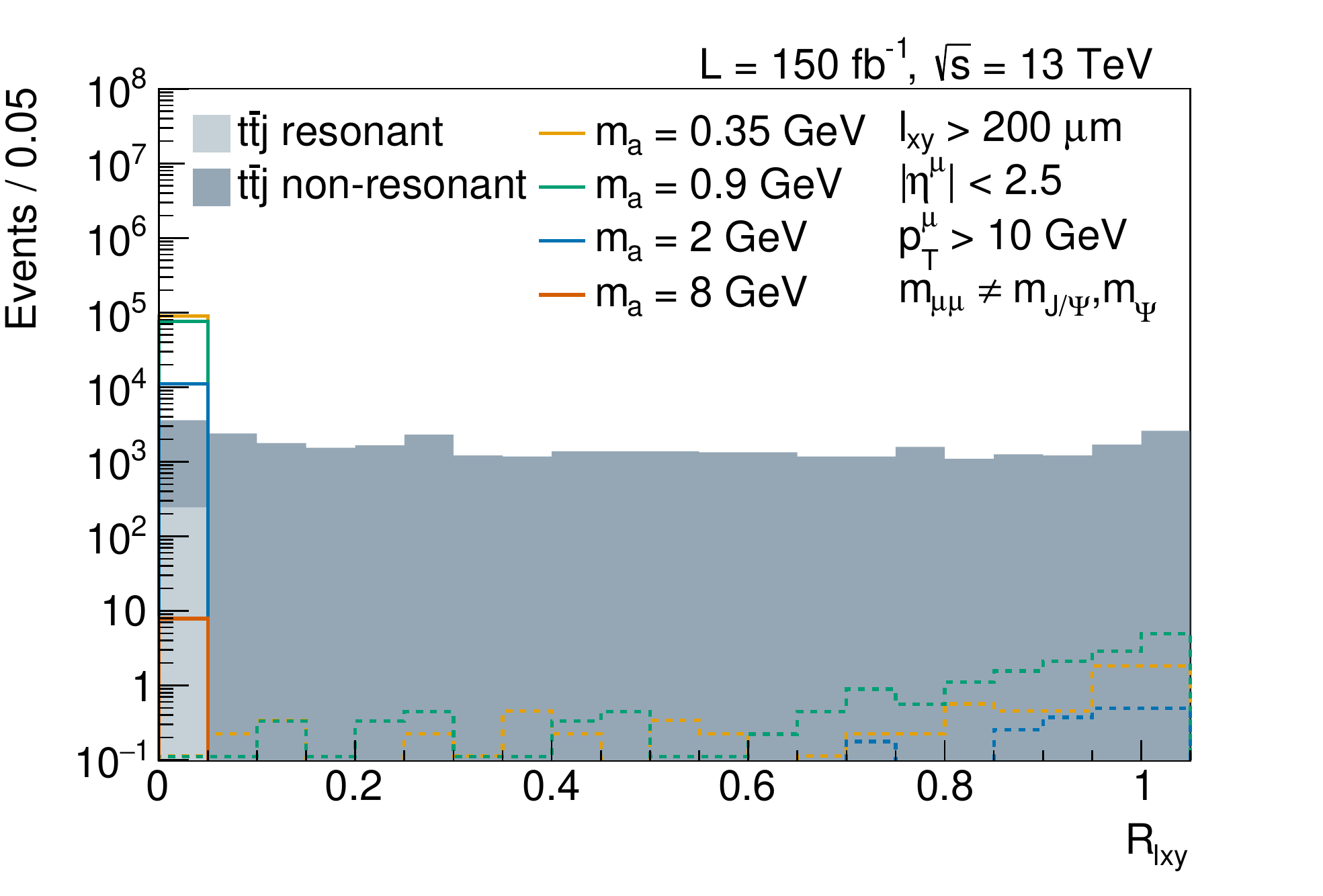}
\caption{\label{fig:di-muon-deltalxy-ratio}The ratio of the transverse displacement, $R_{lxy}$, between the muons of the pre-selected di-muons, for various ALP signal mass benchmarks (colored lines) and background from $t\bar{t}j$ events (gray regions). The distribution is shown after the full event selection (upper right corner), excluding the $R_{lxy}$ criterion.}
\end{figure}

The ratio of the transverse displacement between the muons, $R_{lxy}$, is displayed in Fig.~\ref{fig:di-muon-deltalxy-ratio}. As can be seen, the signal and resonant $t\bar{t}j$ background accumulate in the first bin, while the non-resonant background is nearly flat in $R_{lxy}$. The ratio $R_{lxy}$ is therefore an efficient selection variable to reduce the number of events with muons originating from different production points. To account for detector resolution effects affecting the reconstruction of the di-muons, we pick events for which the di-muon with the smallest value of $R_{lxy}$ passes a relatively loose selection criterion of $R_{lxy} < 0.05$.

 Table~\ref{tab:signal-yields} shows the cumulative efficiency after each applied criterion in the event selection, in order from top to bottom, and the number of events passing the event selection, for four benchmark signals. Similarly, the cumulative background efficiency after each selection stage and the number of passing events are given in Table~\ref{tab:background-yields}. The few remaining $t\bar{t}Z^{(\ast)}$ events after the selection mainly include muons from meson decays. The event selection effectively reduces the overall number of background events, while keeping the number of signal events stable. 
 
\begin{figure}
    \centering
    \includegraphics[width=.85\textwidth]{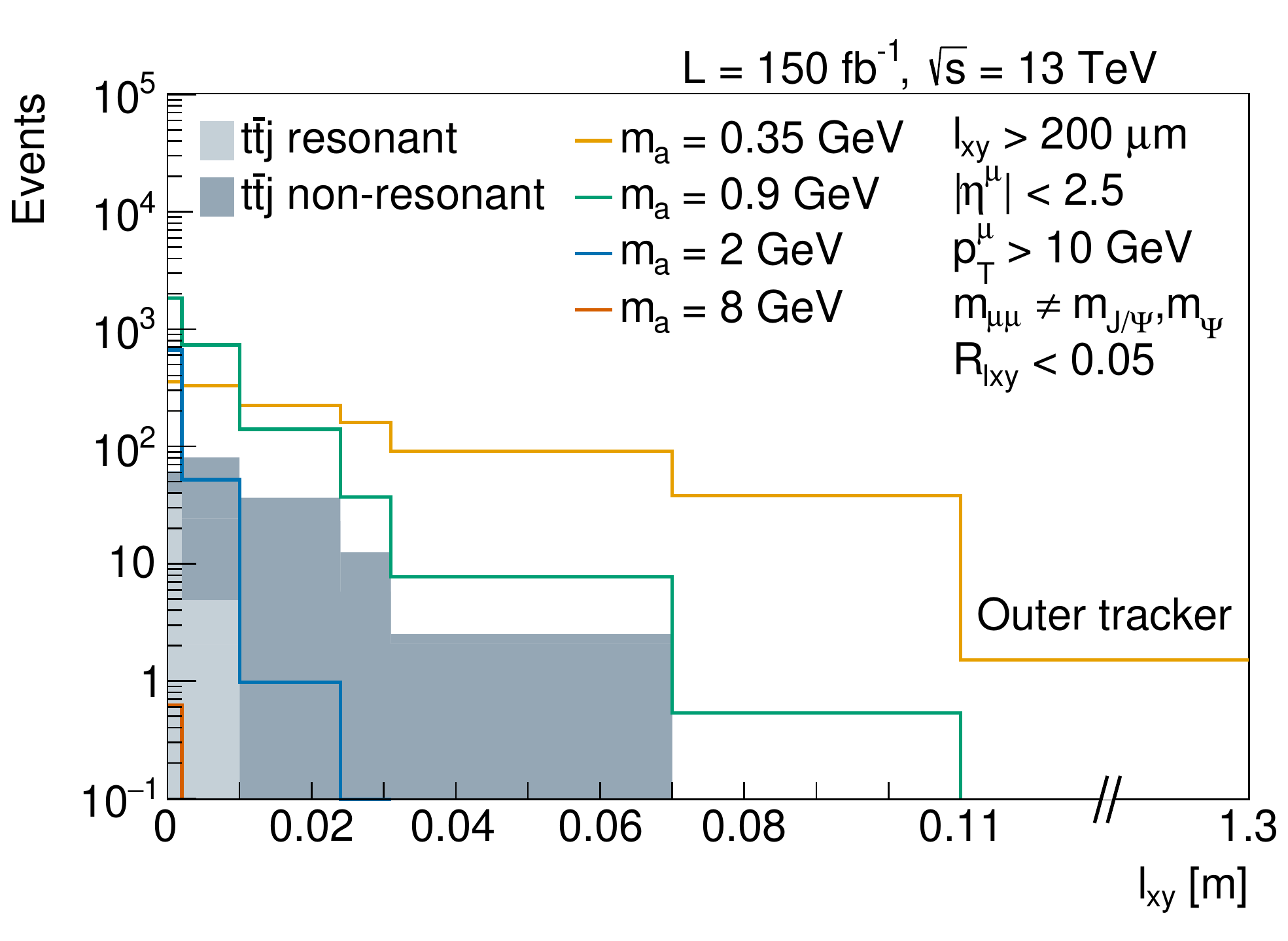}
    \caption{\label{fig:lxy_inner_tracker}The transverse displacement $l_{xy}$ of the least displaced muon after event selection, for various ALP signal benchmarks (colored lines) and background from $t\bar{t}j$ events (gray regions). Displaced decays within the pixel tracker (in 6 bins) and the outer tracker are shown, with the binning used in the analysis.}
\end{figure}

\begin{table}[tbp]
    \centering 
    \resizebox{\textwidth}{!}{ 
    \begin{tabular}{lcccccc} \hline \hline
       Signal efficiency & $m_a = 0.35$ GeV & $m_a = 0.9$ GeV & $m_a = 2$ GeV & $m_a = 8$ GeV  \\
        \hline
        Pre-selection & (8.92 $\pm$ 0.01)$\times$$10^{-1}$ & (7.94 $\pm$ 0.01)$\times$$10^{-1}$ & (6.40 $\pm$ 0.01)$\times$$10^{-1}$ & (7.25 $\pm$ 0.03)$\times$$10^{-2}$ \\
        $p_T^\mu > 10$ GeV & (7.99 $\pm$ 0.01)$\times$$10^{-1}$& (6.79 $\pm$ 0.01)$\times$$10^{-1}$ & (5.58 $\pm$ 0.01)$\times$$10^{-1}$ & (6.87 $\pm$ 0.03)$\times$$10^{-2}$ \\
        $m_{\mu\bar{\mu}} \neq m_{J/\Psi}, m_{\Psi(2S)}$ & (7.99 $\pm$ 0.01)$\times$$10^{-1}$ & (6.79 $\pm$ 0.01)$\times$$10^{-1}$ & (5.58 $\pm$ 0.01)$\times$$10^{-1}$ & (6.86 $\pm$ 0.03)$\times$$10^{-2}$  \\
        $R_{lxy} < 0.05$ & (7.99 $\pm$ 0.01)$\times$$10^{-1}$ & (6.79 $\pm$ 0.01)$\times$$10^{-1}$ & (5.58 $\pm$ 0.01)$\times$$10^{-1}$& (6.86 $\pm$ 0.03)$\times$$10^{-2}$ \\  
        \hline 
        \hline
        Events passing pre-selection & 19793 $\pm$ 21 & 17697 $\pm$ 20 & 2516 $\pm$ 3 & 1.66 $\pm$ 0.01 \\
        Events passing signal selection & 17740 $\pm$ 20 & 15116 $\pm$ 18 & 2193 $\pm$ 3 & 1.57 $\pm$ 0.01 \\
        \hline 
        \hline
    \end{tabular}} 
    \caption{\label{tab:signal-yields}The cumulative selection efficiency and the number of events after the selection for four ALP signal benchmark masses, for an integrated luminosity $\mathcal{L} = 150\,$fb$^{-1}$ and $\sqrt{s} = 13\,$TeV. The listed uncertainties are statistical.}
\end{table}
%
\begin{table}[tbp]
\centering 
    \resizebox{.8\textwidth}{!}{ 
    \begin{tabular}{lcc} \hline \hline
       Background efficiency & $t\bar{t}j$  & $t\bar{t}Z^{(\ast)}$  \\
        \hline
        Pre-selection & (2.55  $\pm$ 0.05)$\times$$10^{-4}$  & (1.89 $\pm$ 0.04)$\times$$10^{-4}$ \\
        $p_T^\mu > 10$ GeV & (7.4 $\pm$ 0.2)$\times$$10^{-5}$  & (9.4 $\pm$ 0.3)$\times$$10^{-5}$  \\
        $m_{\mu\bar{\mu}} \neq m_{J/\Psi}, m_{\Psi(2S)}$ & (6.8 $\pm$ 0.2)$\times$$10^{-5}$  & (5.8 $\pm$ 0.2)$\times$$10^{-5}$  \\
        $R_{lxy} < 0.05$ & (7.1 $\pm$ 0.8)$\times$$10^{-6}$  & (4.9 $\pm$ 0.7)$\times$$10^{-6}$  \\
        \hline 
        \hline
        Events passing pre-selection & 25917 $\pm$ 458 & 0.59 $\pm$ 0.01 \\
        Events passing signal selection & 721 $\pm$ 76 & 0.015 $\pm$ 0.002 \\
        \hline 
        \hline
    \end{tabular}} 
    \caption{\label{tab:background-yields}The cumulative selection efficiency and the number of events after the selection for the background, for an integrated luminosity $\mathcal{L} = 150\,$fb$^{-1}$ and $\sqrt{s} = 13\,$TeV. The listed uncertainties are statistical.}  
\end{table}
\indent To analyze the signal sensitivity in terms of the decay length of the ALPs, we separate the event samples into several bins of the transverse displacement $l_{xy}$. For this analysis, we concentrate on the tracker region and define displacement bins up to $l_{xy} = 1.3\,$m, with the bins for the pixel tracker defined as in Ref.~\cite{CMS:2021sch}. The signal and background $l_{xy}$ distributions of the least displaced muon for events passing all selections are shown in Fig.~\ref{fig:lxy_inner_tracker}.

\begin{figure}[t!]
\centering
    \includegraphics[width=.85\textwidth]{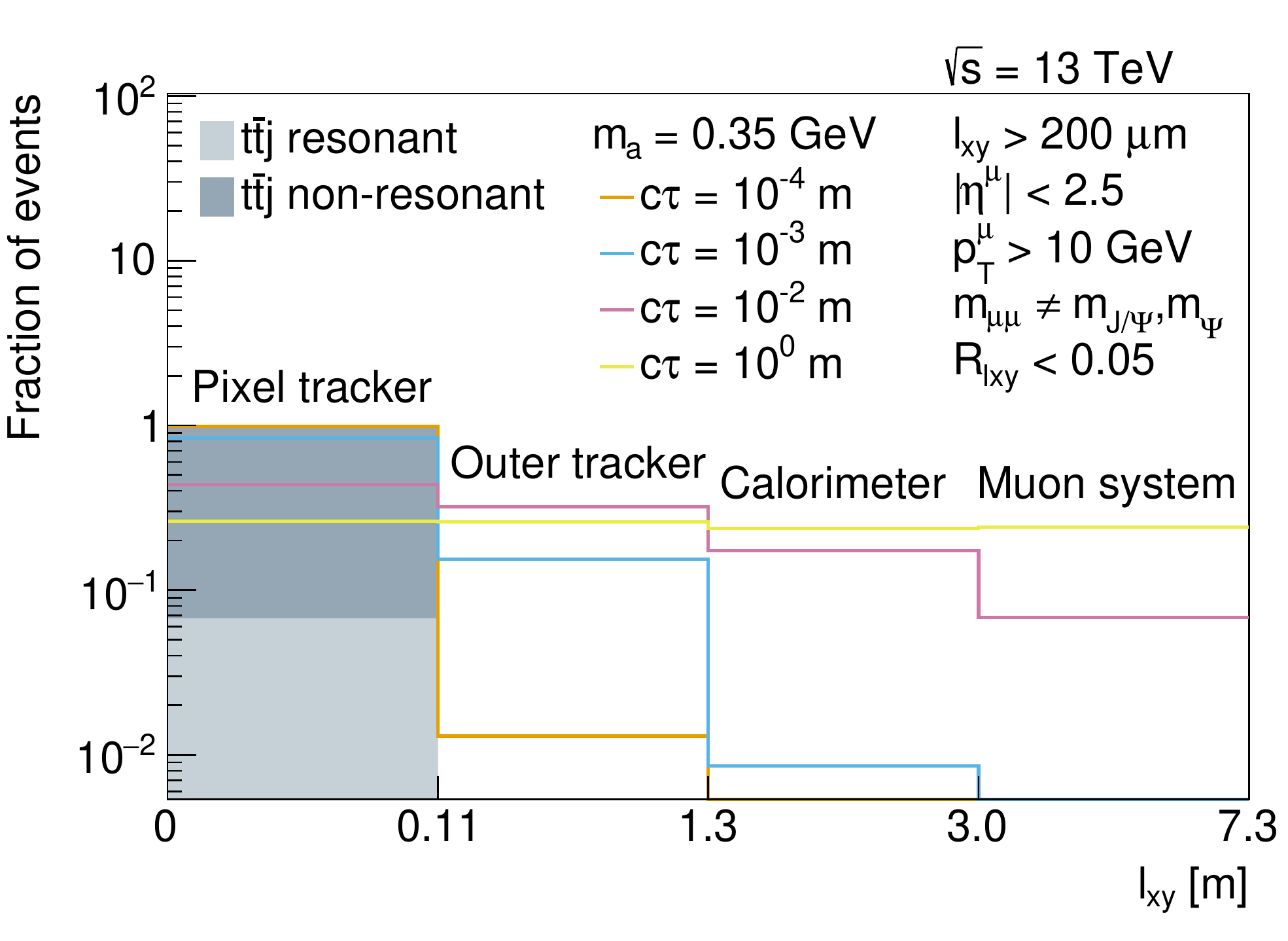}
    \caption{\label{fig:lxy_all}The transverse displacement $l_{xy}$ of the least displaced muon after the event selection, for the $t\bar t j$ background (gray regions) and ALP signal benchmarks with fixed $m_a = 0.35\,$GeV and varying proper lifetime $c\tau$ (colored lines). Events are normalized to unity.}
\end{figure}

In the last step, we generalize the top scenario and treat the lifetime of the ALP as an independent parameter. To this end, we explicitly enforce c$\tau_{a}$ for several values in the range $10^{-4}\,$ -- $10^{0}\,$m. The corresponding $l_{xy}$ distributions for $m_a = 0.35$~GeV are shown in Fig.~\ref{fig:lxy_all}. In this case, the distributions are binned in four large detector regions, approximately based on the CMS detector geometry~\cite{CMS:2008xjf}. In the top scenario, the ALPs decay predominantly within the tracker region (see Fig.~\ref{fig:lxy_inner_tracker}), and the background is contained within the pixel tracker. In an experiment, we would expect some small amount of instrumental background also in the outer tracker, calorimeters, and the muon system. However, we verified that a few background events at higher $l_{xy}$ do not significantly affect the sensitivity of this study. For ALPs and, more generally, for top-associated LLPs with longer lifetimes, searches for displaced decays in the calorimeters and in the muon system can further increase the sensitivity.

\section{Predictions for LHC Run 2 and HL-LHC}
\label{sec:predictions}
After the selection procedure outlined in Sec.~\ref{sec:selection}, we predict the expected sensitivity of the LHC and HL-LHC to top-associated displaced vertices. In Sec.~\ref{sec:alp-results}, we discuss our results for ALPs in the top scenario, which is described by two parameters, $m_a$ and $|c_{tt}|/f_a$. In Sec.~\ref{sec:general-results}, we generalize these results to pseudo-scalars with an arbitrary production rate and lifetime. We compare the predicted sensitivity with existing searches and indirect probes for similar scenarios in Sec.~\ref{sec:comparison}.

To calculate the expected limits, we use the {\tt combine} tool~\cite{CMS-NOTE-2011-005}, which implements the frequentist {\tt AsymptoticLimits} method~\cite{Cowan_2011}. Limits were calculated for ALP mass points (including all 15 benchmarks used in this analysis) ranging from $0.3$ to 10~GeV. Between mass points, we use a linear interpolation.

The uncertainty in the luminosity measurement in CMS is around $2-3\%$ \cite{CMS:LUM-21-001}. However, in order to accommodate other uncertainties affecting both signal and background in the same way (such as displaced vertex reconstruction resolution, or di-muon mass resolution), we assume a conservative overall systematic uncertainty of 10\% in the signal efficiency and 30\% in the background efficiency when computing expected limits.

Note that all of the results are presented assuming $\sqrt{s} = 13\,$TeV, although a center-of-mass energy of 14$\,$TeV is expected at the HL-LHC. Thus, the results presented here for the HL-LHC are conservative, considering the increase in the cross section expected at higher $\sqrt{s}$.

\subsection{ALP benchmark scenario}\label{sec:alp-results}
To calculate the expected upper limits on the cross section $\sigma(pp\to t\bar{t}a)\times \mathcal{B}(a\to \mu\bar{\mu})$ and on the ALP-top coupling $|c_{tt}|/f_a$, we use the events that pass the event selection shown in Tables~\ref{tab:signal-yields} and \ref{tab:background-yields}. We categorize these events based on the displacement of the reconstructed vertex. These categories include six bins for the pixel tracker and one bin for the outer tracker (see Fig.~\ref{fig:lxy_inner_tracker}).

\begin{figure}[t!]
\centering
    \includegraphics[width=0.49\linewidth]{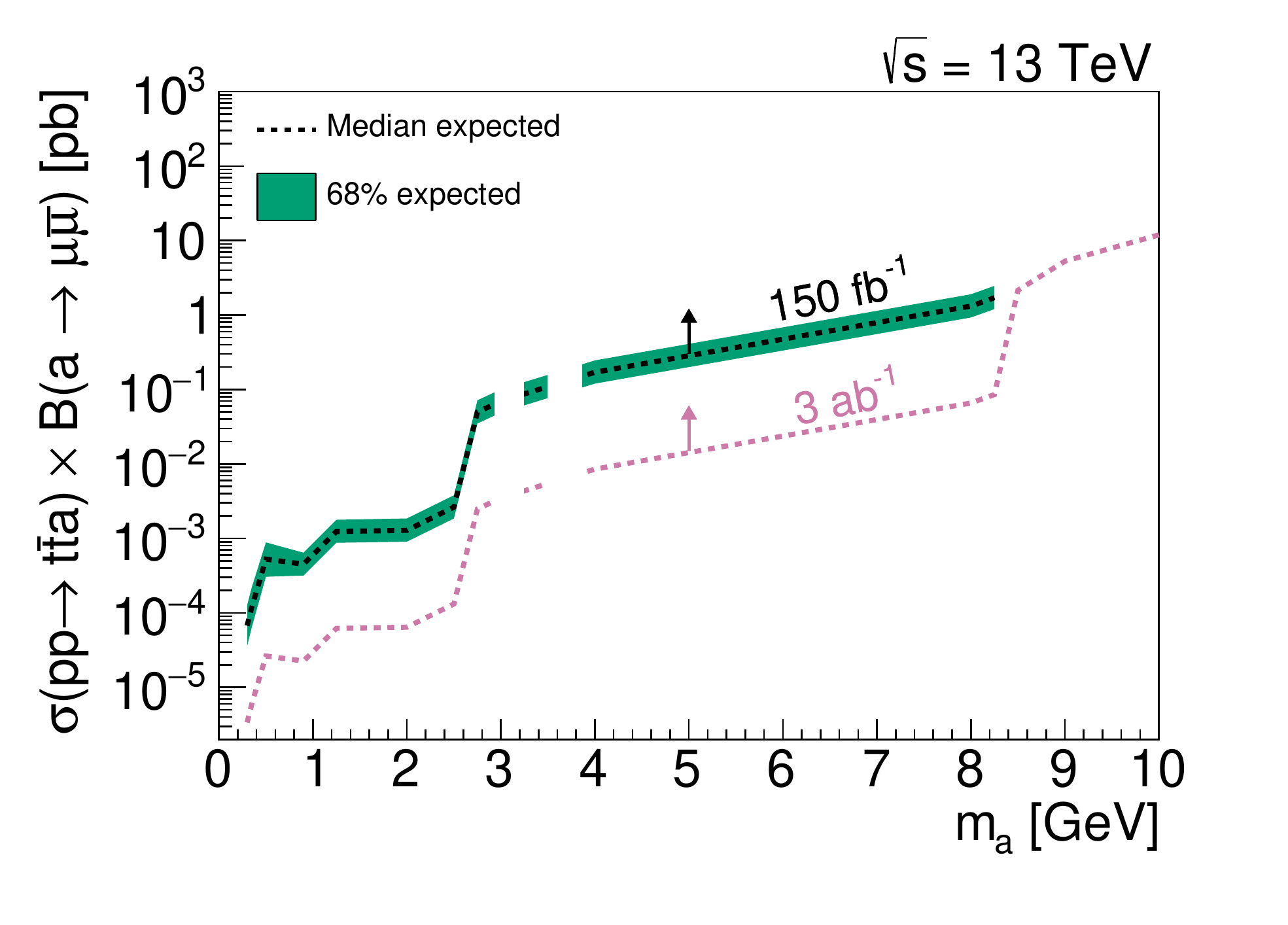}
    \includegraphics[width=0.49\linewidth]{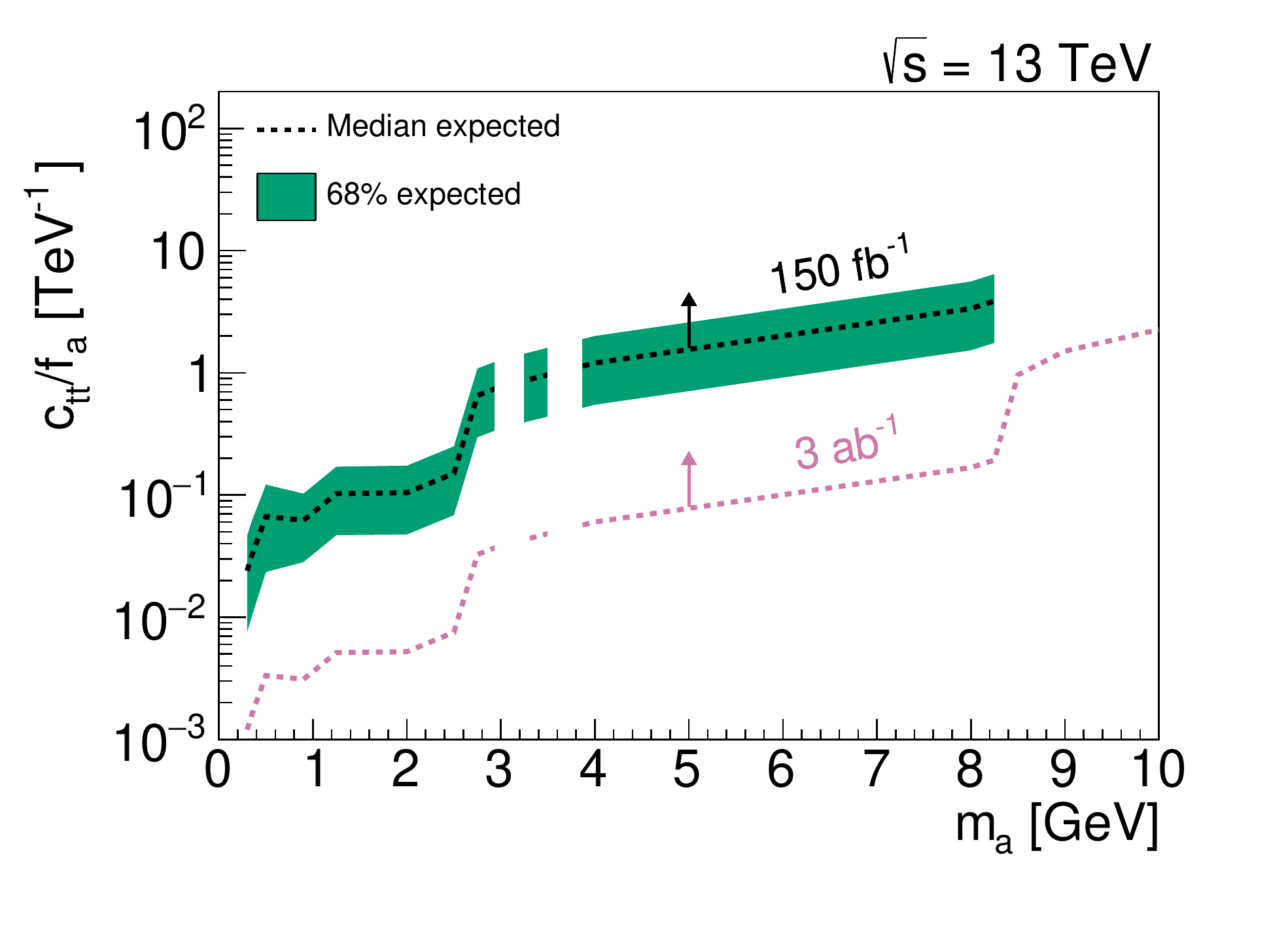}
\caption{\label{fig:limit-xsec_coupling}Expected 95\% CL upper limits on the cross section times branching ratio (left) and the ALP-top coupling $|c_{tt}(\Lambda)|/f_a$ (right) in the top scenario, as a function of the ALP mass $m_a$. The signal rate and lifetime of the ALP are set by the two model parameters $m_a$ and $c_{tt}(\Lambda)/f_a$, with $f_a = 1\,$TeV and $\Lambda = 4\pi f_a$. Predictions are shown for an integrated luminosity of 150 fb$^{-1}$ (3 ab$^{-1}$), corresponding to the LHC Run~2 (HL-LHC). ALP masses around the $J/\Psi$ and $\Psi(2S)$ resonances are removed by the event selection, as indicated by the interruptions of the curves. For ALP masses above $8.25\,$GeV, the expected event rate with $150\,$fb$^{-1}$ is close to zero and these points are therefore excluded.}
\end{figure}

The results are shown in Fig.~\ref{fig:limit-xsec_coupling}, where we display the expected 95\% confidence level (CL) upper limits on $\sigma(pp\to t\bar{t}a)\times \mathcal{B}(a\to \mu\bar{\mu})$ (left panel) and on $|c_{tt}|/f_a$ (right panel), as a function of the ALP mass.
 The predicted limits correspond to an integrated luminosity of $150\,$fb$^{-1}$ collected at the LHC during Run 2. To obtain projections for the HL-LHC, we simply upscale the data luminosity to $3\,$ab$^{-1}$. The sensitivity is highest for small ALP masses and drops quickly at larger masses. This tendency can be understood intuitively: light ALPs have longer lifetimes and statistically decay in regions further away from the production point, which are less affected by SM background (see Fig.~\ref{fig:lxy_inner_tracker}). Furthermore, for light ALPs the branching ratio into muons is larger than for heavier ALPs, where other decay channels are kinematically accessible. This results in much lower di-muon event rates for heavy ALPs (see Table~\ref{tab:signal-yields}). The sensitivity to heavy ALPs could be enhanced by up to 2 orders of magnitude in certain cases by searching for other final states, for instance for di-taus or multi-hadron final states. The exact gain from including non-muon final states would require a dedicated study. A dedicated search for promptly decaying pseudo-scalars produced in association with a top-antitop pair (see Sec.~\ref{sec:comparison}) could also be sensitive to ALPs with masses above the GeV scale, since most of them decay very close to the production point.
 
As indicated in Sec.~\ref{sec:signal}, throughout our analysis, we have assumed that we can identify muons coming from the top quark decay with an efficiency of $\epsilon_{t\bar{t}}=100\%$. We have also assumed that top quarks can be fully reconstructed regardless of their decay chain. In reality, both signal and background will be affected by the top reconstruction efficiency. We expect that a limited efficiency will scale down the number of signal and background events in a similar way, because they both contain $t\bar{t}$, and low-mass ALPs do not affect their kinematics. This will reduce the overall sensitivity in Fig.~\ref{fig:limit-xsec_coupling} and shift the limits to a higher effective cross section $\sigma\times \mathcal{B}/\epsilon_{t\bar{t}}$. In a similar way, the choice of the top decays in an analysis will scale the reach proportionally to the top branching ratio. For instance, if the overall top-pair reconstruction efficiency was at the level of 10\%, the limits on $\sigma\times \mathcal{B}$ in Fig.~\ref{fig:limit-xsec_coupling} would shift upwards by about an order of magnitude.


\subsection{General pseudo-scalars with top couplings}\label{sec:general-results}
Using the predicted event rates for arbitrary lifetimes of the ALP, see Fig.~\ref{fig:lxy_all}, we interpret our analysis for general pseudo-scalar particles with couplings to top quarks.\footnote{The shift symmetry $a\to a + c$ of the ALP Lagrangian~\eqref{eq:lagrangian} does not affect the event kinematics in $pp\to t\bar{t}a$ production. An ALP with generic couplings and lifetime has therefore the same phenomenology as any pseudo-scalar that does not feature a shift symmetry.} Unlike in the top scenario, the cross section and lifetime of the pseudo-scalar are not correlated.

\begin{figure}[t!]
\centering
\includegraphics[width=.85\textwidth]{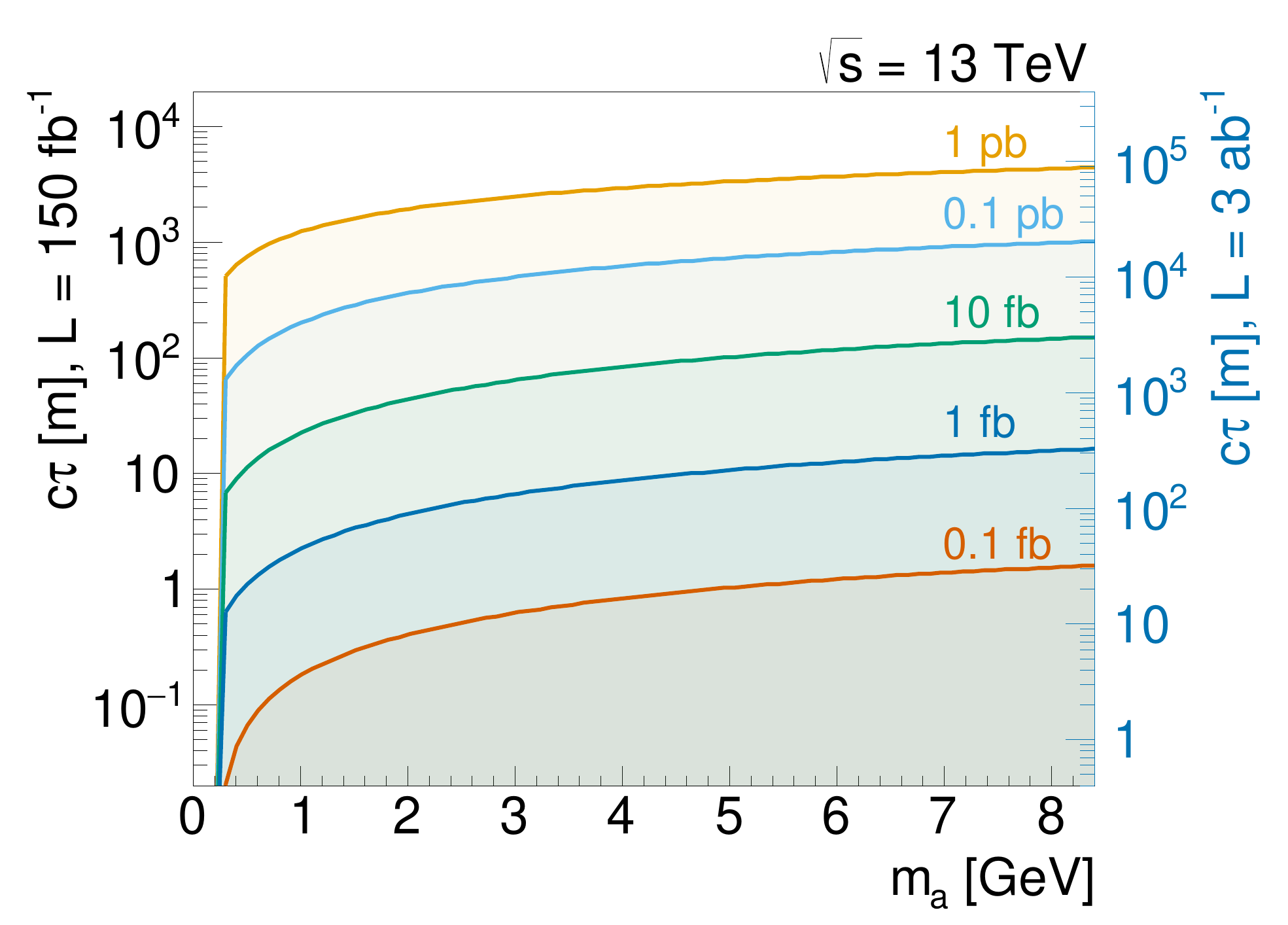}
\caption{\label{fig:limit-lifetime}The expected sensitivity for a pseudo-scalar with proper lifetime $c\tau_{a}$ and mass $m_a$, for fixed values of the signal cross section $\sigma(pp\to t\bar{t}a)\times\mathcal{B}(a \to \mu\bar{\mu})$. The shaded areas can be excluded at 95\% CL. The left (right) vertical axis corresponds to an integrated luminosity of 150 fb$^{-1}$ (3 ab$^{-1}$).}
\end{figure}

In Figure~\ref{fig:limit-lifetime}, we show the expected 95\% CL upper limits on the proper decay length of the pseudo-scalar, $c\tau_{a}$, as a function of its mass, $m_a$, for different assumptions on $\sigma(pp\to t\bar{t}a)\times\mathcal{B}(a \to \mu\bar{\mu})$. With data corresponding to $150\,$fb$^{-1}$ of integrated luminosity and cross sections around $1\,$fb, pseudo-scalars with proper lifetimes up to around 1--$10\,$m can be probed in the considered mass range. For longer lifetimes, a large fraction of particles decays outside of the detector and escapes the analysis acceptance. At the HL-LHC with an integrated luminosity of $3\,$ab$^{-1}$, the reach can be extended by more than an order of magnitude in the proper lifetime. For very short lifetimes of $c\tau_{a} \lesssim 100\,\mu$m, the signal is difficult to distinguish from the meson-induced background. As a consequence, at lifetimes below those displayed in Fig.~\ref{fig:limit-lifetime}, the signal sensitivity drops.

\subsection{Comparison with existing searches}\label{sec:comparison}
In searches for long-lived particles, the sensitivity depends on both the signal cross section and the decay length. A comparison with searches with prompt or invisible final states is therefore not straightforward, but possible within certain limitations. Here we attempt to compare our predictions for top-associated displaced di-muons with (proposed) searches for similar ALP scenarios at the LHC and other experiments. Most of these searches are complementary in mass and/or lifetime to this study.

\paragraph{Promptly decaying pseudo-scalars in top-antitop production} To obtain an optimal sensitivity across a large range of lifetimes, LHC searches for displaced vertices should be combined with searches for promptly decaying resonances and searches with missing energy. The ATLAS and CMS collaborations have performed searches for prompt di-muons from pseudo-scalars produced via $pp\to t\bar t a,\,a \to \mu\bar{\mu}$~\cite{ATLAS:2023ofo,CMS:2019lwf,CMS:2022arx}. The ATLAS analysis~\cite{ATLAS:2023ofo} selects prompt di-muons with invariant masses $15\,$GeV$\,< m_{\mu\mu} < 72\,$GeV from final states with multi-leptons and jets. For pseudo-scalars with masses in this range, the search is sensitive to cross sections $\sigma(pp\to t\bar{t}a)\times \mathcal{B}(a\to \mu\bar{\mu})$ of a few femtobarns with $139\,$fb$^{-1}$ of data. CMS achieves a similar sensitivity within the same mass range~\cite{CMS:2022arx}. The reach is comparable with our predictions for displaced di-muons from pseudo-scalars with $2 m_\mu < m_a < 10\,$GeV and proper decay lengths $1\,\text{m} < c\tau_{a} < 10\,$m in $150\,$fb$^{-1}$ of data (see Fig.~\ref{fig:limit-lifetime}). For pseudo-scalars with decay lengths $c\tau \lesssim 1\,$m, a larger sensitivity in cross section can be reached with displaced-vertex searches, provided that the displacement is large enough to be resolved.

\paragraph{Invisible pseudo-scalars in top-pair production} Searches for top quarks and missing energy in $t\bar{t}$ production~\cite{Haisch:2016gry} and single-top production~\cite{Pinna:2017tay,Plehn:2017bys} have been proposed to probe (pseudo-)scalars with top couplings. The corresponding searches by ATLAS and CMS~\cite{CMS:2015zwg,CMS:2017dcx,CMS:2017qxu,CMS:2019zzl,CMS:2018ysw,ATLAS:2018cjd,ATLAS:2014bba,ATLAS:2018nda,ATLAS:2017hoo,ATLAS:2014dbf} target heavy particles well above $10\,$GeV. In Ref.~\cite{Esser:2023fdo}, the authors have re-interpreted an ATLAS search for top-pairs and missing energy~\cite{ATLAS:2021hza} for detector-stable ALPs in a broader mass range (see also Ref.~\cite{Ebadi:2019gij}). They find an upper bound on the ALP-top coupling, $|c_{tt}|/f_a < 1.8/$TeV at $95\%\,$CL, provided that the ALP escapes detection. In the top scenario, this condition applies for ALPs with masses well below the di-muon threshold. Missing energy and displaced vertex searches therefore probe complementary mass regimes of ALPs with top couplings. Alternatively, the ALP-top coupling can be probed indirectly through NLO effects in $t\bar{t}$ production~\cite{Galda:2021hbr,Esser:2023fdo}. The sensitivity is comparable to top-associated missing energy searches.

\paragraph{Displaced ALPs from top decays} In Ref.~\cite{Carmona:2022jid}, the authors have proposed a search for displaced ALPs produced from top decays via flavor-changing top-charm and top-up couplings at the LHC. The analysis focuses on the process $pp\to t\bar{t},\,t\to a q$, with subsequent hadronic ALP decays. The signature consists of a trackless jet that could be detected in the hadronic calorimeter or the muon spectrometer. Similar to our strategy, the authors suggest using the top quark to trigger on these types of events. For $t\bar{t}$ events, this analysis is sensitive to top quark branching ratios $\mathcal{B}(t\to aq) \lesssim 10^{-4}$ for hadronically decaying ALPs with masses above $1\,$GeV. A recast of a CMS search for flavor-changing top-quark couplings in single top events yields a sensitivity to $\mathcal{B}(t\to aq) \approx 10^{-3}$. The sensitivity to flavor-changing ALP couplings is comparable to our prediction for flavor-diagonal ALP couplings. In this way, both analyses complement each other in probing the parameter space of the ALP effective Lagrangian.

\paragraph{Inclusive di-muon searches} An alternative to top-associated di-muons is inclusive searches for (displaced) di-muon resonances. Several displaced di-muon searches have probed resonances with masses well above $10\,$GeV~\cite{CMS:2022qej,ATLAS:2018rjc}. To access di-muons with smaller invariant masses, a dedicated search for displaced di-muon vertices using a high-rate data stream (``scouting'', ``turbo'', or ``trigger-level'' analysis) has been developed~\cite{CMS:2021sch}. On the other hand, such scouting searches store reduced detector information, thus deteriorating the tracking performance and limiting the range of observable displacements. Moreover, di-muon searches based on scouting data do not allow one to reconstruct additional top quarks. Therefore, we expect that inclusive di-muon searches are less sensitive to top-philic ALPs or other light particles, due to a larger amount of background. Notice, however, that a direct comparison of inclusive and top-associated di-muon searches is largely model-dependent because the relative event rates depend strongly on the relevant couplings of the particles.

\paragraph{ALP signatures with virtual top quarks} Indirect bounds on the ALP-top coupling can be obtained by exploring loop-induced ALP couplings to other particles through virtual tops~\cite{Bauer:2020jbp,Bonilla:2021ufe}. As mentioned in Sec.~\ref{sec:alps}, an ALP-top coupling at the cutoff scale $\Lambda = 4\pi f_a$ of the ALP Effective Field Theory induces ALP couplings to all other SM particles at lower energies. At the LHC, effective ALP couplings to photons, weak gauge bosons, and the Higgs can be probed in, for instance, di-boson production and $h\to Za$ decays~\cite{Bauer:2017ris,Gavela:2019cmq,CMS:2016369}. In the top scenario, the derived indirect bounds on the ALP-top coupling range around $|c_{tt}|/f_a \lesssim 10-100/$TeV at $95\%\,$CL~\cite{Esser:2023fdo}. With the current experimental status, the sensitivity to the ALP coupling to virtual tops is weaker than in searches with resonant tops.

A very sensitive probe of ALPs below the GeV scale is meson decays like $B \to K a$, induced through top-quark loops. For flavor-universal couplings $c_{f\!f}/f_a$, the sensitivity of ALP decays into muons is stronger than in top-associated ALP production~\cite{Ferber:2022rsf}. In the top scenario, however, the decay length of the ALP is comparably larger, so that searches for displaced di-muons from meson decays lose sensitivity. Proposed searches for $B \to K a$ decays with missing energy at Belle II~\cite{Ferber:2022rsf} offer an interesting alternative to probe the top scenario. A dedicated comparison of flavor and high-energy observables in this scenario would give valuable information on future prospects.

 As described in this section, there exist a large variety of searches for ALPs. However, all of them probe different regions of phase space (including lower or higher masses), prompt decays, or do not require top-antitop pair in an event, which results in overwhelming backgrounds. Therefore, we believe that the analysis strategy proposed in this work is complementary to existing studies and would nicely fit in the landscape of searches for ALPs.

\section{Summary and outlook}
\label{sec:summary}
This work aimed to establish the discovery potential of top-associated displaced vertices at the LHC. To this end, we have performed an analysis of ALPs produced in association with a top-antitop pair and decaying into a displaced di-muon. In a benchmark study, the so-called top scenario, we have assumed that the ALP couples only to top quarks at high energies, so that the effective top coupling $c_{tt}/f_a$ and the mass $m_a$ of the ALP are the only model parameters.

Based on simulation, we have designed a dedicated selection procedure to discriminate between the top-associated di-muon signal and the dominant background from $t\bar{t}Z^{(\ast)}$ and $t\bar{t}j$ production. For a fixed ALP coupling $c_{tt}/f_a = 1$/TeV, we have quantified step by step how the event selection efficiently reduces the background while preserving most of the signal. A crucial discriminator is the displacement of the reconstructed di-muon vertex, which suppresses much of the background from meson decays with observable displacements.

Assuming 100\% effective identification of muons from top quarks and 100\% trigger efficiency for top quarks, as well as including all top quark decay modes, we derive the expected reach of the LHC for top-philic ALPs in Fig.~\ref{fig:limit-xsec_coupling} and for generalized long-lived pseudo-scalars in Fig.~\ref{fig:limit-lifetime}. In the top scenario, the (HL-)LHC is sensitive to ALP-top couplings as small as $c_{tt}/f_a \approx 0.03 (0.002/$TeV) for ALP masses $2 m_\mu < m_a \lesssim 2\,$GeV with $150\,$fb$^{-1}$ ($3\,$ab$^{-1}$). More generally, top-philic pseudo-scalars with proper lifetimes up to $20\,$m ($300\,$m) and cross sections around $1\,$fb can be probed at the (HL-)LHC. 

Our results suggest that searches for top-associated displaced di-muons bridge the current sensitivity gap in the lifetime between searches with prompt di-muons (short lifetimes) and those with missing energy (detector-stable particles). In many scenarios, the lifetime of a particle is anti-correlated with its mass. This suggests that top-associated displaced di-muons will lead the sensitivity to top-philic ALPs within the range $2 m_\mu < m_a \lesssim 2\,$GeV at the LHC. From an experimental perspective, displaced vertex searches in association with reconstructed top quarks offer new options for triggering. This opens new opportunities to extend the discovery potential of the LHC to feebly interacting particles well below the weak scale.

\acknowledgments

 JA, FB, JN, and LR acknowledge support from DESY (Hamburg, Germany), a member of the Helmholtz Association HGF, and support by the Deutsche Forschungsgemeinschaft (DFG, German Research Foundation) under Germany’s Excellence Strategy -- EXC 2121 "Quantum Universe" -- 390833306. RS acknowledges support of the DFG through the research training group Particle Physics Beyond the Standard Model (GRK 1940). The research of SB and SW has been supported by the DFG
under grant no. 396021762–TRR 257.


\newpage

\bibliographystyle{JHEP}
\bibliography{ttalp}

\end{document}